\documentclass[prb,reprint,showpacs,showkeys]{revtex4-1}
\usepackage{graphicx}
\usepackage{amsmath,amssymb}
\usepackage{bm}
\newcommand{\br}[1]{\textbf{\textrm{#1}}}
\newcommand{\eb}{\varepsilon_\mathrm{B}}
\newcommand{\ee}{\varepsilon_0}
\newcommand{\cc}{\mathrm{c.c.}}
\newcommand{\ib}{\textit{ibid.~}}
\newcommand{\sech}{\mathrm{sech}}
\newcommand{\im}{\mathrm{Im~}}
\newcommand{\re}{\mathrm{Re~}}
\newcommand{\rmx}{\mathrm{x}}
\newcommand{\rmf}{\mathrm{MF}}
\newcommand{\rfm}{\mathrm{FM}}

\begin{document}
\title{Modulational instability and solitons in excitonic semiconductor waveguides}
\author{Oleksii A. Smyrnov}
\email{oleksii.smyrnov@mpl.mpg.de}
\author{Fabio Biancalana}
\affiliation{Nonlinear Photonic Nanostructures Group, Max Planck Institute for the Science of Light, G\"{u}nther-Scharowsky
Str. 1/26, 91058 Erlangen, Germany}
\author{Stefan Malzer}
\affiliation{Department of Applied Physics, Friedrich-Alexander University of Erlangen-Nuremberg, Staudtstr. 7/A3, 91058 Erlangen, Germany}
\date{\today}
\begin{abstract}
Nonlinear light propagation in a single-mode micron-size waveguide made of semiconducting excitonic material has been theoretically studied in terms of exciton-polaritons by using an analysis based on macroscopic fields. When a light pulse is spectrally centered in the vicinity of the ground-state Wannier exciton resonance, it interacts with the medium nonlinearly. This optical cubic nonlinearity is caused by the repulsive exciton-exciton interactions in the semiconductor, and at resonance it is orders of magnitude larger than the Kerr nonlinearity (e.g., in silica). We demonstrate that a very strong and unconventional modulational instability takes place, which has not been previously reported. After reducing the problem to a single nonlinear Schr\"{o}dinger-like equation, we also explore the formation of solitary waves both inside and outside the polaritonic gap and find evidence of spectral broadening. A realistic physical model of the excitonic waveguide structure is proposed.
\end{abstract}
\maketitle

\section{Introduction\label{s1}}
The nonlinear phenomena of modulational instability (MI) and solitons formation make the basis of modern nonlinear fiber optics, where the Kerr (cubic) nonlinearities are mainly used.~\cite{agrawal} MI consists in the exponential ("parametric") growth of equally spaced spectral sidebands from a continuous wave pump, and is equivalent to the degenerate four-wave mixing.~\cite{agrawal} It is a particularly useful phenomenon in microstructured fibers, where it allows the generation of coherent light at frequencies where lasers are not easily available.~\cite{wadsworth} The MI instability is strictly connected to the existence of solitons, or solitary waves, i.e. localized short pulses that are invariant in the propagation, and the shape of which is determined by the specific nonlinearity and the dispersive features of fiber.~\cite{agrawal,hasegawa} Plane waves tend to become unstable due to the fact that the actual "normal modes" of the fiber are nonlinear optical solitons.~\cite{agrawal,mollenauer} A particularly useful phenomenon connected to solitons is the so-called supercontinuum generation, which is the sudden and explosive spectral broadening of an input energetic short pulse in the fiber.~\cite{dudley}

However, optical fibers are not the only media where MI and solitons can be observed and used. The general investigations of optical nonlinear effects in dielectrics and semiconductors also have history of several decades,~\cite{akimoto,knorr,ostr,knorr2,talanina93,talanina,giessen,kamch} and recently many applied researches have been started to investigate the potential of the solitary wave sector of semiconductor materials near excitonic resonances, for example, in semiconductor microcavities or semiconductor gratings.~\cite{dmitry,bianex} The dominant optical cubic nonlinearity for the polarization field in excitonic materials, which we investigate here, is caused by the exciton-exciton Coulomb repulsive interactions and essentially differs from the cubic Kerr nonlinearity for the electric field (e.g., in silica). The excitonic nonlinearity is orders of magnitude larger than the Kerr one, but it also possesses a resonant nature - as we will see in the following, it substantially weakens when the detuning of the central frequency of the pulse from the exciton resonance increases. The first difference allows one to expect the observation of MI and solitons formation not on the kilometer (e.g., like in silica fibers) but on micrometer scale and for lower pulse intensities. This opens up the possibility of on-chip integration of optical devices for nonlinear frequency conversion based on waveguides made of semiconducting excitonic materials.

In this work we theoretically study all the mentioned optical nonlinear properties of such excitonic waveguides. In Sec.~\ref{s2} on the basis of known nonlinear differential equation for the complex envelope of macroscopic medium polarization field~\cite{ostr} and Maxwell's equations for the light pulse we derived a general set of equations, which govern the light propagation in the excitonic waveguide. Further, we applied this set of equations to investigate the instability of long light pulses propagating in the excitonic waveguide with respect to small perturbations and without any approximations discovered a very strong modulational instability of the pulse spectrum with enormous gain (Sec.~\ref{s3}). This is completely confirmed by the direct numerical simulation of the pulse spectrum evolution using the general set of equations. Next, we reduced the general set of equations to a single nonlinear differential equation for the polarization field envelope, which in its turn under the slowly varying envelope approximation (SVEA) was reduced to the nonlinear Schr\"{o}dinger equation with an exact and complete set of coefficients. Within the so-called "tail analysis" formalism~\cite{akimoto} all the parameters of corresponding solitonic solution of this equation have been obtained for light pulses spectrally centered both outside and even inside the polaritonic gap, where the propagation of linear plane waves is forbidden, while solitons can nevertheless form and propagate if the incident pulse intensity is large enough (Sec.~\ref{s4}).
In this way we have substantially generalized previous theories in the existing literature \cite{akimoto,kamch} without any additional approximations. These analytical derivations are also supported by direct numerical simulations using the general set of equations under SVEA. Finally, the obtained results allowed us to propose a realistic physical model of the excitonic waveguide and discuss the experimental conditions for the observation of above-mentioned effects using GaAs as the representative excitonic material (Sec.~\ref{s5}).

\section{Nonlinear equation for polarization\label{s2}}
In Ref.~\onlinecite{ostr} it was shown that the set of semiconductor Bloch equations for the excitonic medium in the low excitation regime can be reduced to a single nonlinear differential equation for a macroscopic complex envelope $P(z,t)$ of the medium polarization $\mathcal{P}(\br{r},t)=\left\{P(z,t)M(\br{r}_\perp)\exp[ikz-i\omega t]+\cc\right\}/2$ coupled to the external electric field $\mathcal{E}(\br{r},t)=\left\{E(z,t)F(\br{r}_\perp)\exp[ikz-i\omega t]+\cc\right\}/2$, where $z$ is the longitudinal coordinate and $\br{r}_\perp$ are the transverse coordinates. Here we assume the existence of a fundamental guided mode, thus allowing a factorization of the dimensionless transverse modal distributions $M$ and $F$ from their respective envelopes. Maintaining the most essential nonlinear terms, after the separation of variables this equation can be written in the following form:
\begin{equation}\label{polar_ini}
\begin{split}
i\partial_t P + (\Delta\omega_\rmx &+ i\gamma_\rmx)P - \alpha \Gamma_1|P|^2 P\\
&+ \tilde{a}\eb E(\Gamma_\rmf-\beta \Gamma_2|P|^2) = 0,
\end{split}
\end{equation}
where $\Delta\omega_\rmx=\omega-\omega_\rmx$ is the detuning of the pulse spectrum central frequency $\omega$ from the 1s-exciton resonance $\omega_\rmx$, $\gamma_\rmx$ is the exciton damping parameter introduced here phenomenologically, $\alpha$ and $\beta$ - nonlinear coefficients, which are completely defined by the microscopic (quantum) properties of the excitonic medium (within the jellium two-band exciton model one has, as in Ref.~\onlinecite{ostr}: $\alpha=26\omega_\mathrm{b}/3$ and $\beta=7$, where $\omega_\mathrm{b}$ is the exciton binding frequency), $\eb$ is the bulk background dielectric constant and $\tilde{a}$ is the photon-exciton coupling parameter. It is well-known that the photon-exciton coupled state, the so-called polariton, shows a region of forbidden frequencies in its dispersion relation - the polaritonic gap, which width is given by $\tilde{a}$. Parameters $\Gamma$ depend on the relation between $F$ and $M$: $\Gamma_1\equiv\int|M|^4d\br{r}_\perp/\int|M|^2d\br{r}_\perp$, $\Gamma_2\equiv\int|M|^2\bar{M}Fd\br{r}_\perp/\int|M|^2d\br{r}_\perp$, $\Gamma_\rmf\equiv\int\bar{M}Fd\br{r}_\perp/\int|M|^2d\br{r}_\perp$. The modal distributions $F$ and $M$ are localized functions,~\cite{agrawal} therefore $\Gamma_1\lesssim1$, $\Gamma_2\sim1$, and $\Gamma_\rmf\gtrsim1$, while $\Gamma_\rfm\lesssim1$ as the polarization field modal distribution $M$ is completely determined by that of the electric field $F$. The cubic term in Eq.~\eqref{polar_ini} originates from the repulsive Coulomb exciton-exciton interactions in the semiconductor, while the square term is responsible for the phase-space filling effect (also called the Pauli blocking) when the density of excitons is comparable to the Mott density (for which $\beta|P|^2\thicksim 1$), what leads to a decoupling of the external electric field from the polarization field.

In the low excitation regime considered in Ref.~\onlinecite{ostr} for exciton densities far below the Mott density, one can approximate Eq.~\eqref{polar_ini} as:
\begin{equation}\label{polar_main}
i\partial_t P + (\Delta\omega_\rmx + i\gamma_\rmx)P-\alpha \Gamma_1|P|^2P+\tilde{a}\Gamma_\rmf\eb E=0.
\end{equation}
We perform the following analytical derivations and numerical simulations on the basis of Eq.~\eqref{polar_main} coupled to Maxwell's equations for the electric and magnetic field envelopes of the incident pulse, $\mathcal{H}(\br{r},t)=\left\{H(z,t)F(\br{r}_\perp)\exp[ikz-i\omega t]+\cc\right\}/2$. In the most general dimensionless form this set of equations is given by:
\begin{equation}\label{sys_ini}
\left\{\begin{array}{l}
 \partial_x\eta=(i\omega'-\partial_T)(\psi+\Gamma_\rfm\lambda\varphi)-ik'\eta,\\
 \partial_x\psi=(i\omega'-\partial_T)\eta-ik'\psi,\\
 \partial_T\varphi=i\left(\left[i\gamma'_\rmx+\Delta\omega'_\rmx-|\varphi|^2\right]\varphi+\Gamma_\rmf\psi\right),\\
      \end{array}\right.
\end{equation}
where the following redefinitions and scalings have been applied: $\psi=E/E_0$, $\varphi=P/P_0$, $\eta=H/H_0$, $x=z/z_0$, $T=t/t_0$, $\omega'=\omega t_0$, $\gamma'_\rmx=\gamma_\rmx t_0$, $\Delta\omega'_\rmx=\Delta\omega_\rmx t_0$, $\lambda=\tilde{a}t_0$, $k'=kz_0$, and $P_0=1/\sqrt{\Gamma_1\alpha t_0}$, $E_0=P_0/(\eb \tilde{a}t_0)$, $H_0=nE_0$, $z_0=ct_0/n$, $c$ is the velocity of light in vacuum, $n=\sqrt{\eb}$ is the non-resonant background refractive index, and $t_0$ is an arbitrary time scaling parameter, which will be properly chosen in the following.

\section{Modulational instability analysis\label{s3}}
We now use system~\eqref{sys_ini} to analyze the linear stability of a long pulse (ideally a continuous wave) propagating in the excitonic optical waveguide with respect to small perturbations. This is the MI analysis, see also Ref.~\onlinecite{agrawal}. By substituting the perturbed field and polarization envelopes $\{\psi;\eta;\varphi\}(x,T)=(\{\psi_0;\eta_0;\varphi_0\}+\{a;g;p\}(x,T))\exp[iqx]$ into system~\eqref{sys_ini}, setting the small perturbations as $\{a;g;p\}(x,T)=\{a_1;g_1;p_1\}\exp[i\kappa x - i\delta T]+\{a_2;g_2;p_2\}\exp[i\delta T-i\kappa x]$ and imposing the solvability condition of system~\eqref{sys_ini} one can obtain within the first order perturbation theory~\cite{agrawal} the dispersion relation $\kappa(\delta)$ for perturbations:
\begin{widetext}
\begin{equation}\label{mi_matrix}
\left|
  \begin{array}{cccc}
    (\omega'+\delta)^2-(k'+q+\kappa)^2 & 0 & \Gamma_\rfm\lambda(\omega'+\delta)^2 & 0 \\
    0 & (\omega'-\delta)^2-(k'+q-\kappa)^2 & 0 & \Gamma_\rfm\lambda(\omega'-\delta)^2 \\
    \Gamma_\rmf & 0 & \Delta\omega'_\rmx+\delta-2\varphi^2_0 & -\varphi^2_0 \\
    0 & \Gamma_\rmf & -\varphi^2_0 & \Delta\omega'_\rmx-\delta-2\varphi^2_0 \\
  \end{array}
\right|=0,
\end{equation}
\end{widetext}
where $\kappa$ is the perturbation wavenumber, $\delta$ is the detuning from the central pump frequency, and $q=\omega'[1+\lambda \Gamma_\rmf \Gamma_\rfm/(\varphi^2_0-\Delta\omega'_\rmx)]^{1/2}-k'$. Those frequency regions where $\im{\kappa(\delta)}<0$ correspond to the exponential growth of perturbations and this defines the MI gain spectrum $G(\delta)=2\max\left|\im{\kappa(\delta)}\right|$. Among all the solutions of Eq.~\eqref{mi_matrix}, which is of the fourth order in $\kappa$, we select, for a given $\delta$, the one with the maximum absolute value of the gain.

\begin{figure}[t]
  \includegraphics[height=5cm]{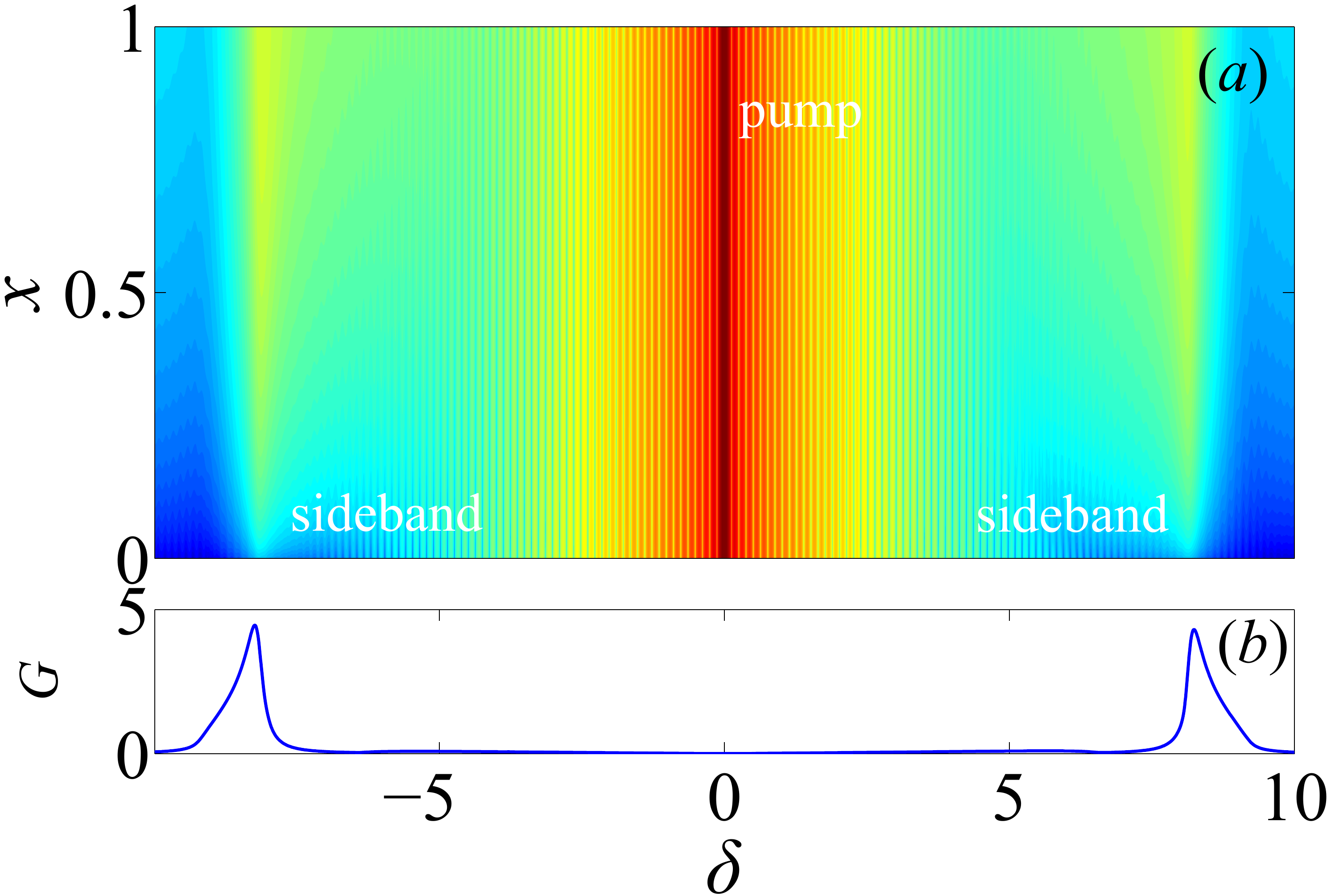}
  \caption{(a) Numerically obtained spectral evolution of a long pulse propagating along the $x$-axis (according to~\eqref{sys_ini}, the initial pulse is $\psi(x=0,T)=\psi_0\exp\left[-(2T/T_\mathrm{w})^{2m}\right]$, ~$T_\mathrm{w}=50, ~m=20, ~\psi_0=10.2$, and the other parameters are $\Delta\omega'_\rmx=-1.1, ~\gamma'_\rmx\simeq0.11$). (b) Corresponding analytical prediction on the MI gain $G(\delta)$ peaks, given by~\eqref{mi_matrix}. $\omega'$ is positioned at the origin.}\label{fig_mi}
\end{figure}

To confirm the above analytical results we performed direct numerical simulations of the propagation of long pulses in the excitonic waveguide by using system~\eqref{sys_ini} without any approximations and obtained the spectrum evolution (see Fig.~\ref{fig_mi}(a)) in complete accordance with the analytical predictions on the MI gain maxima positions (see Fig.~\ref{fig_mi}(b)).

A fourth order Runge-Kutta algorithm for the magnetic, electric and polarization fields has been employed with the following initial conditions: $\psi(x=0,T)$ is a wide super-Gaussian pulse, $\eta(x=0, T)=\pm\psi(x=0, T)$ (for respectively forward and backward propagation) and $\varphi(x=0,T=0)=0$, which means that the medium is not polarized before the field arrival. We choose $t_0=1/\tilde{a}$ (for GaAs this is $\sim8$~ps) as natural temporal scale of the system. Here we set $k'=\omega'$ and for the simulations removed the large number $\omega'\sim\omega_\rmx/\tilde{a}$ from~\eqref{sys_ini} by multiplication of the two first equations by $\lambda/\omega'$ and space rescaling $z_0\rightarrow z_0\lambda/\omega'\sim c/(n\omega)\sim0.04~\mu\mathrm{m}$, which is the general physical space scaling of the whole developed model (from now on GaAs is selected as a representative material, for numbers see Sec.~\ref{s5}). With this scaling it can be seen from Fig.~\ref{fig_mi} that the MI gain at its maxima reaches values $\sim100~\mu\mathrm{m}^{-1}$, which is enormous in comparison with that for silica optical fibers possessing nonlinear Kerr effect with the MI gain $\sim10~\mathrm{km}^{-1}$. This example clearly shows that the nonlinear phenomena, which manifest themselves, e.g., in optical fibers on the km-distances, can be observed in the excitonic waveguides when the incident pulse has propagated for just a few microns.

\begin{figure}[t]
  \includegraphics[height=5cm]{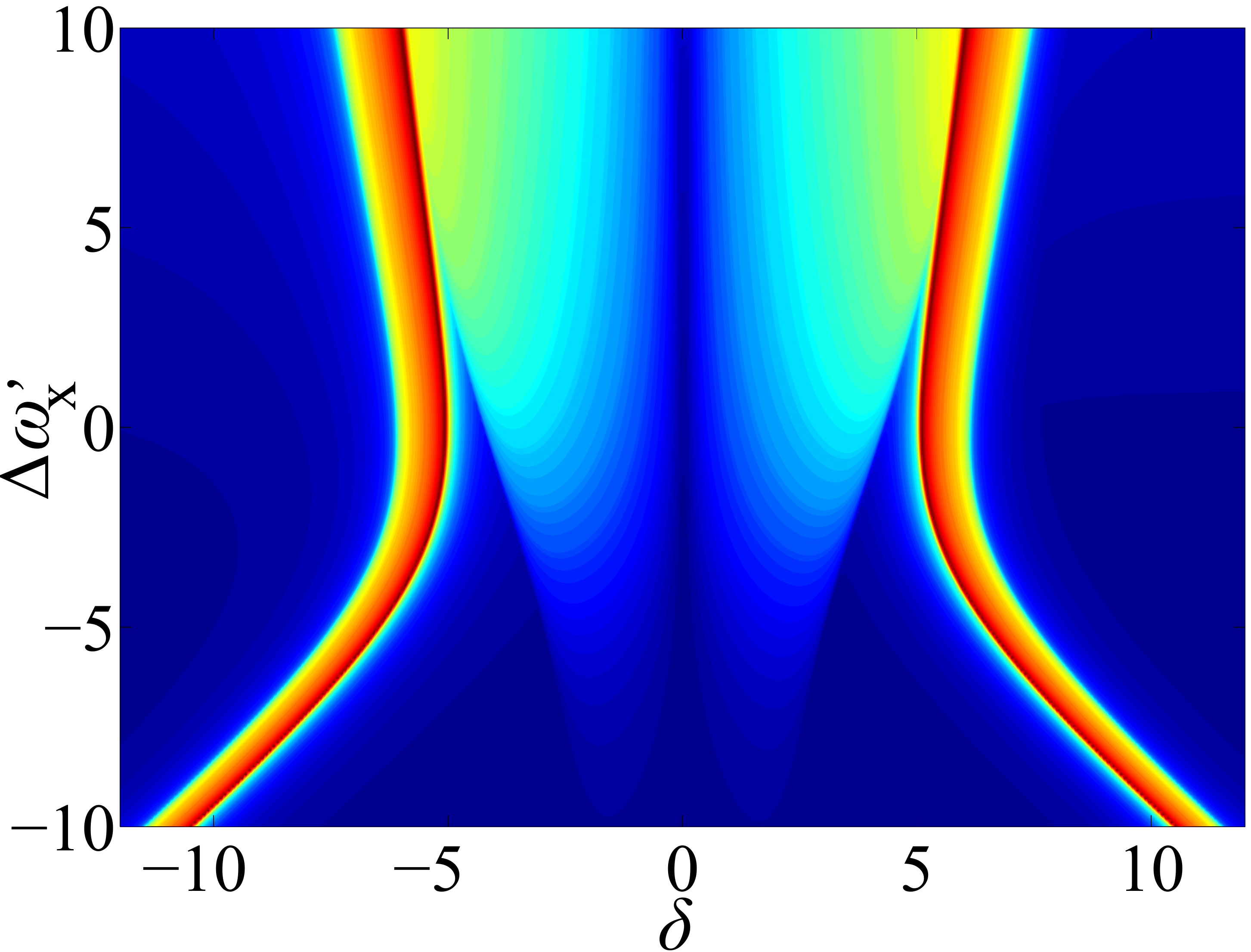}\\
  \caption{Dependence of the MI gain $G(\delta)$ on the frequency detuning $\Delta\omega_\rmx'$ according to \eqref{mi_matrix} for the fixed value of external electric field amplitude $\psi_0=(\varphi_0/\Gamma_\rmf)(\varphi_0^2-\Delta\omega'_\rmx)$. The used parameters are $\psi_0=5, ~\gamma'_\rmx\simeq0.019$.}\label{fig_mi2}
\end{figure}

It should be noted that in our model the excitonic damping $\gamma'_\rmx$ has been naturally taken into account not only in the simulations but also in the analytical results by the redefinition $\Delta\omega'_\rmx\rightarrow\Delta\omega'_\rmx+i\gamma'_\rmx$. The increase of $\gamma'_\rmx$ makes the MI peaks smaller in gain and spectrally wider. The dependence of MI gain on the detuning $\Delta\omega'_\rmx$ with small fixed nonvanishing $\gamma'_\rmx$ is very unconventional and is shown at Fig.~\ref{fig_mi2}. Peaks with unusually large gain appear in the spectrum because of the resonant nature of the excitonic nonlinearity, which, of course, affects the dispersion relation~\eqref{mi_matrix} - complex poles appear in its roots. The positions of these resonant gain peaks shown at Fig.~\ref{fig_mi2} are completely determined by such poles, and the magnitude of the gain is consequently much larger than in non-resonant instabilities. In Fig.~\ref{fig_mi2} also much lower central peaks can be seen, which are almost flat in the scale of Fig.~\ref{fig_mi}(b). These peaks shapes are analogous to those observed in the MI gain spectra of optical fibers in the vicinity of the pump frequency, where only non-resonant Kerr nonlinearity is present.~\cite{agrawal}

\section{Formation of solitary waves\label{s4}}
We have also studied the solitary wave formation under intense and short pulse excitation of the excitonic waveguide in the coherent regime ($\gamma_\rmx=0$). After the change of variables $\xi=x-VT$ (where $V$ is the dimensionless soliton group velocity) system~\eqref{sys_ini} in analogy with Ref.~\onlinecite{stoychev} can be reduced to a single third order nonlinear differential equation for the polarization envelope $\varphi$:
\begin{equation}\label{third}
\begin{split}
i\rho_1\partial_\xi^3\varphi + \rho_2\partial_\xi^2\varphi + &i\rho_3\partial_\xi\varphi - \rho_4\varphi+\sigma_1\partial_\xi^2\left[|\varphi|^2\varphi\right]\\
&+i\sigma_2\partial_\xi\left[|\varphi|^2\varphi\right]+\sigma_3|\varphi|^2\varphi=0,
\end{split}
\end{equation}
where:
\begin{equation*}
\begin{split}
&\rho_1\equiv V(V^2-1),\\
&\rho_2\equiv \lambda_\Gamma V^2-(V^2-1)\Delta\omega'_\rmx-2V(V\omega'-k'),\\
&\rho_3\equiv 2V\omega'\lambda_\Gamma-2\Delta\omega'_\rmx(V\omega'-k')+V(k'^2-\omega'^2),\\
&\rho_4\equiv (k'^2-\omega'^2)\Delta\omega'_\rmx+\lambda_\Gamma\omega'^2,\\
&\sigma_1\equiv V^2-1,~\sigma_2\equiv2(V\omega'-k'),~\sigma_3\equiv k'^2-\omega'^2,
\end{split}
\end{equation*}
and $\lambda_\Gamma=\lambda \Gamma_\rmf \Gamma_\rfm$. Under the slowly varying envelope approximation~\cite{agrawal} (SVEA), because $\rho_{i+1}\gg\rho_i,~i={1,2,3}$, and $\sigma_{i+1}\gg\sigma_i,~i={1,2}$, Eq.~\eqref{third} can be reduced to the nonlinear Schr\"{o}dinger equation:
\begin{equation}\label{nlse}
\rho_2\partial^2_\xi\varphi-\rho_4\varphi+\sigma_3|\varphi|^2\varphi=0,
\end{equation}
where we also set $\rho_3=0$ as the definition of group velocity.~\cite{stoychev} As is well known,~\cite{mihalache} Eq.~\eqref{nlse} has the solitonic solution:
\begin{equation}\label{sech}
\varphi=d~\sech\left[\xi/b\right],
\end{equation}
where $d=\sqrt{2\rho_4/\sigma_3}$, and $b=\sqrt{\rho_2/\rho_4}$. Therefore, the electric field takes the form:
\begin{equation}\label{psech}
\psi=\varphi/\Gamma_\rmf\left(-iV/b~\tanh\left[\xi/b\right]-\Delta\omega'_\rmx+|\varphi|^2\right).
\end{equation}
It is important to note, that here SVEA has been used only after the reduction of the full system~\eqref{sys_ini} to the single third order differential equation, that, unlike Ref.~\onlinecite{talanina}, makes the coefficients in~\eqref{nlse} exact and keeps the most essential cubic term at $\sigma_3$.

In order to explicitly obtain the dispersion relation and other parameters of the propagating pulse we apply the so-called "tail analysis" described, for example, in Ref.~\onlinecite{akimoto}. Starting from the full system~\eqref{sys_ini} and looking for solutions in the form $\{\psi;\phi;\eta\}(\xi)=\{\psi_0;\phi_0;\eta_0\}\exp[-\xi/V]$, which reproduces the exponentially localized shape of a soliton for $\xi\rightarrow\infty$ (at its "tail"), one can linearize the system~\eqref{sys_ini} and finally obtain:
\begin{widetext}
\begin{equation}\label{sol_par}
\begin{split}
  k'=\frac{\omega'\zeta}{n},\qquad V=\frac{\zeta n}{\zeta_1},\qquad &d^2=2\left[\Delta\omega'_\rmx+\frac{\lambda_\Gamma}{\zeta^2/\eb-1}\right],\\
  b^2=\frac{V^2}{\omega'^2}~\frac{\lambda_\Gamma-(1-1/V^2)\Delta\omega'_\rmx-2(\omega'-k'/V)}{(k'^2/\omega'^2-1)\Delta\omega'_\rmx+\lambda_\Gamma},\qquad &\zeta^2=\frac{1}{2}\left[\zeta_2+\left(\zeta^2_2+\frac{4\zeta^2_1}{\omega'^2}\right)^{1/2}\right],\\
  \zeta_1=\eb\left[1+\frac{\lambda_\Gamma\left(\omega'^2-2\omega'\Delta\omega'_\rmx-1\right)}{2\omega'(1+\Delta\omega'^2_\rmx)}\right], \qquad &\zeta_2=\frac{\eb}{1+\Delta\omega'^2_\rmx}\left[\left(1-\frac{1}{\omega'^2}\right)\left(1+\Delta\omega'^2_\rmx-\lambda_\Gamma\Delta\omega'_\rmx\right)-\frac{2\lambda_\Gamma}{\omega'}\right].
  \end{split}
\end{equation}
\end{widetext}
The set of Eqs.~\eqref{sol_par} completely defines all soliton parameters (the dispersion law $k'(\omega')$, the group velocity $V$, the amplitude $d$, the spatial width parameter $b$) and contains just a single free parameter $t_0$, which in this Section is the soliton temporal width. All the other parameters are fully determined by the excitonic medium properties ($\omega_\rmx$, $\tilde{a}$, $\eb$, etc.).

\begin{figure}[t]
  \includegraphics[width=7cm]{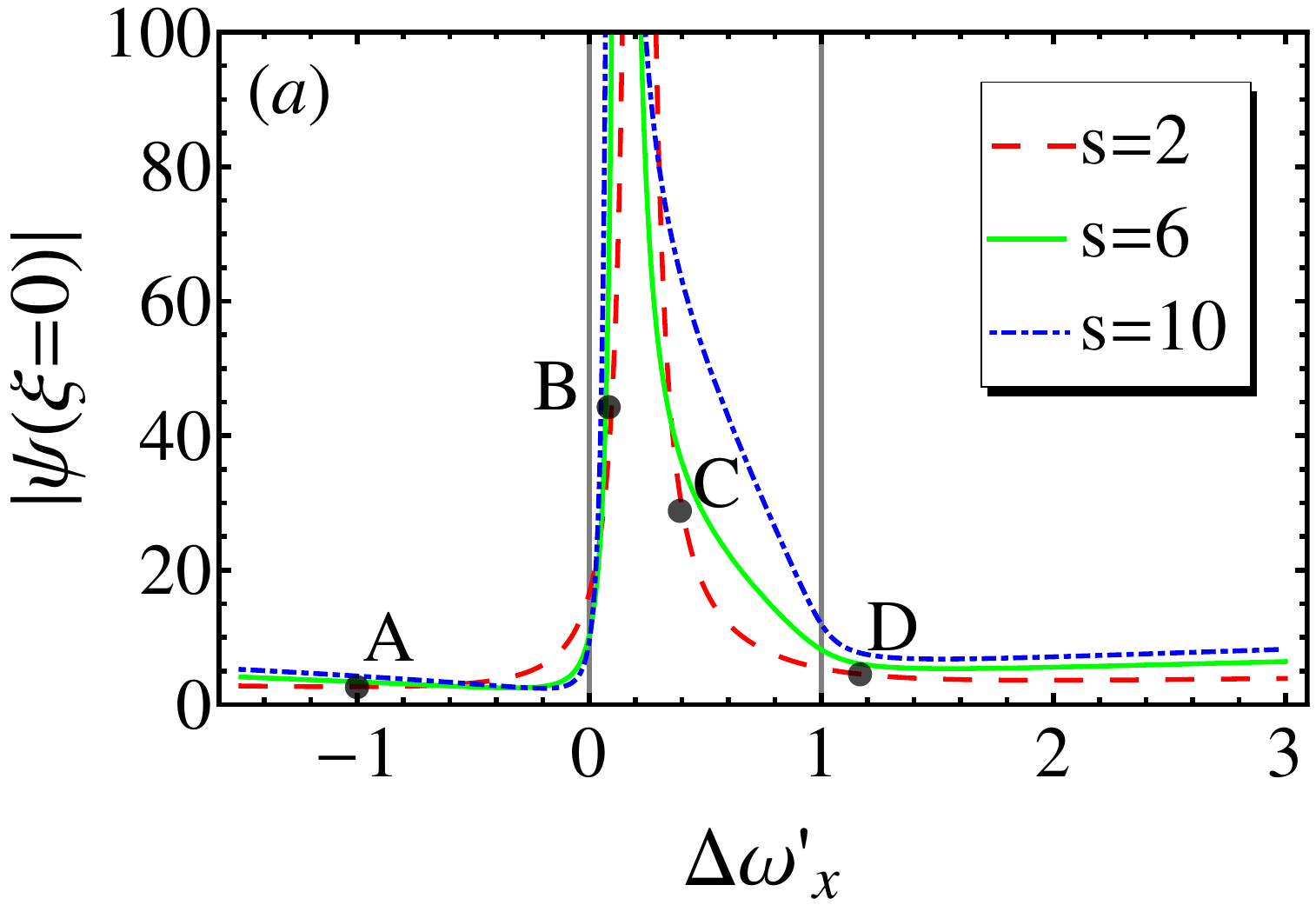}
  \includegraphics[width=7cm]{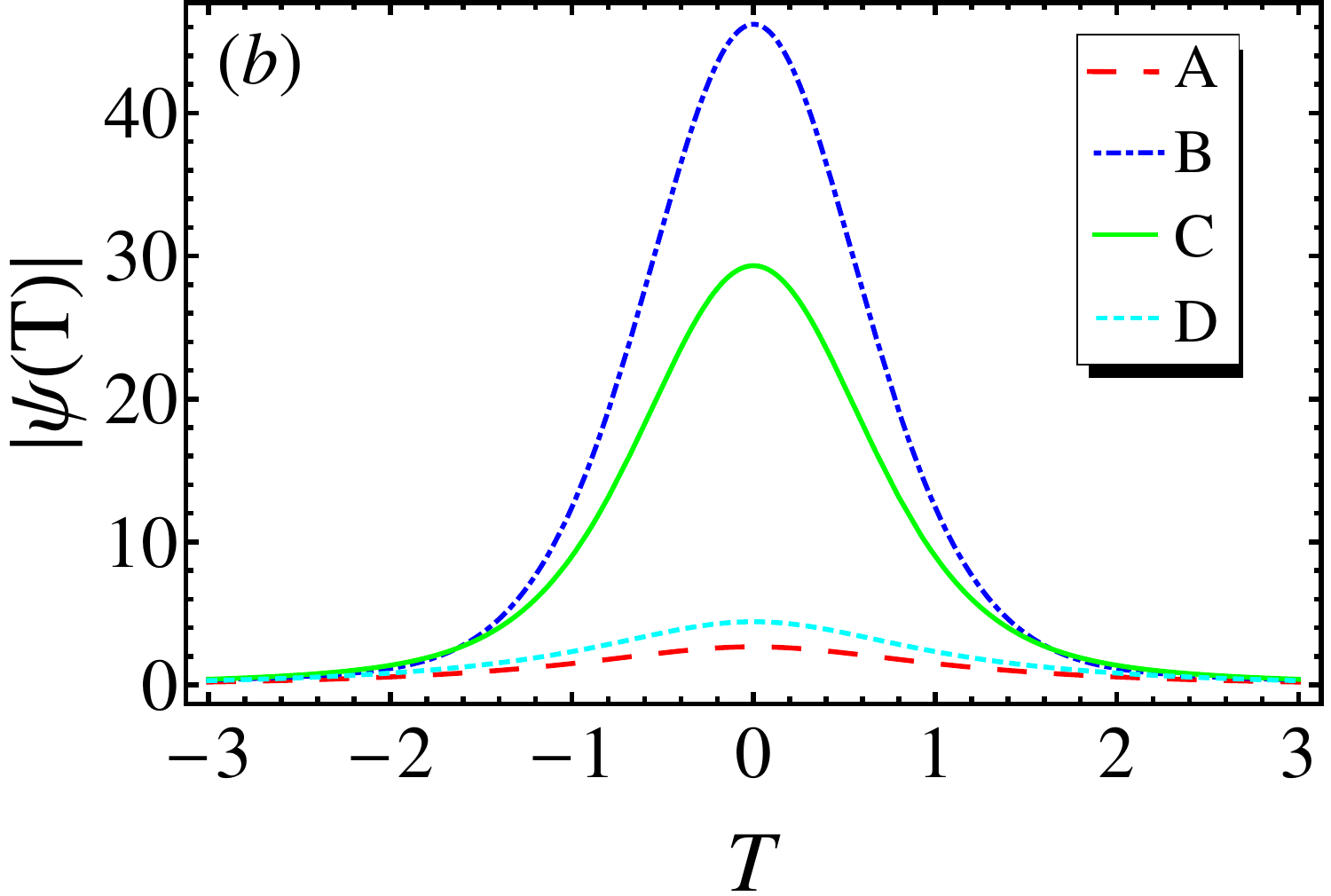}
  \caption{(a) The amplitude of $|\psi|$ pulse soliton as a function of the detuning $\Delta\omega'_\rmx$ (in the units of dimensionless gap $\lambda_\Gamma$) for the different soliton temporal widths $t_0=s/\tilde{a}$ according to~\eqref{sech}-\eqref{sol_par}. The exciton resonance is positioned at the origin, the gap is indicated by the grey lines. (b) The corresponding temporal shapes of solitons both inside and outside the polaritonic gap in the case of $s=2$.}\label{fig_ampl}
\end{figure}

Although the developed model is valid for relatively wide solitons ($t_0\gtrsim1/\tilde{a}$), it generalizes previous theories present in the literature,~\cite{akimoto,kamch} and allows one to calculate all the soliton parameters for the incident pulses spectrally centered both outside and even inside the polaritonic gap, where solitons could also form if the incident pulse intensity is large enough (see Fig.~\ref{fig_ampl}). Previously (e.g., in Refs.~\onlinecite{akimoto,kamch}), in- and out-of-gap solitons were treated separately with some additional approximations. It is clear from Fig.~\ref{fig_ampl} that to form in-gap solitary waves one should input an intensity, which is orders of magnitude higher than in the case of the out-of-gap solitons. The amplitude of the latter grows with the detuning because far from the exciton resonance the nonlinearity is weaker (see Sec.~\ref{s5}) and for the observation of nonlinear effects the input intensity should be increased.

\begin{figure}[t]
  \includegraphics[width=7cm]{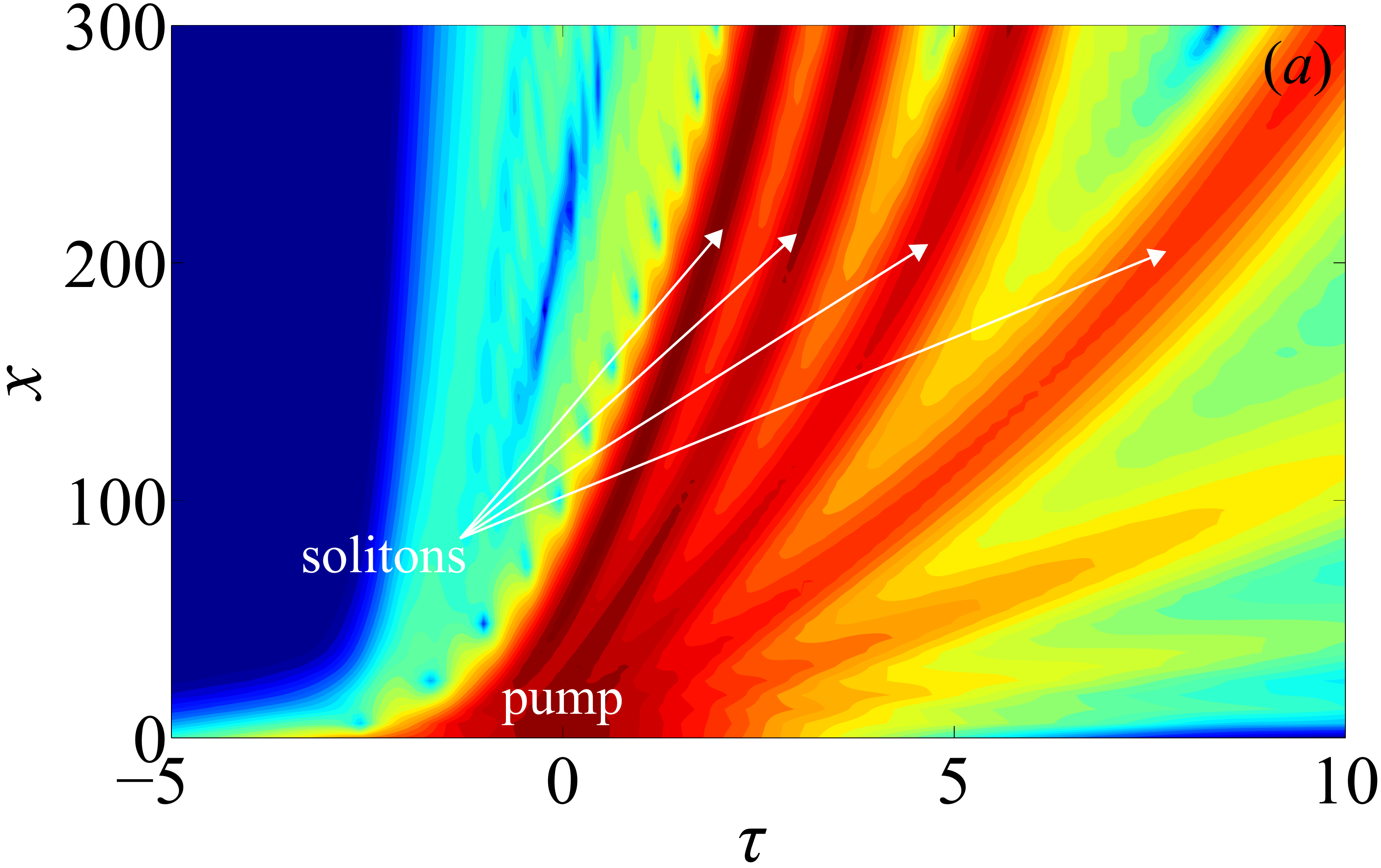}
  \includegraphics[width=7cm]{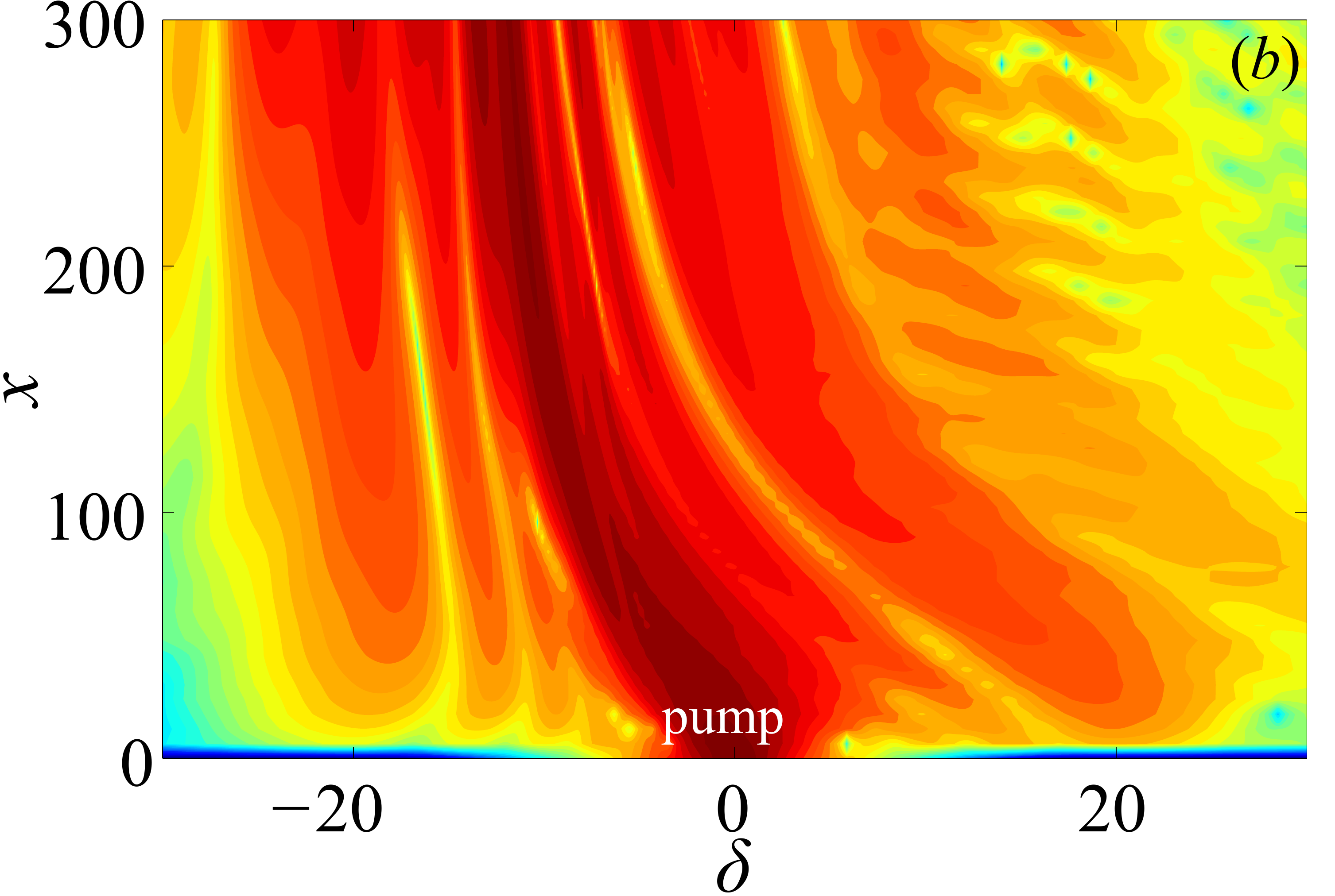}
  \caption{(a) The formation of well separated out-of-gap solitary waves is numerically demonstrated for the short intense sech-shaped $|\psi|$ pulse propagation in the excitonic medium. (b) The evolution of corresponding spectrum with the essential broadening. Calculations are performed by using system~\eqref{sys_svea} and initial pulse $\psi(x=0,\tau)=\psi_0~\sech(\tau), ~\psi_0=20$, while other parameters are $\Delta\omega'_\rmx=1.2, ~\gamma'_\rmx\simeq0.19$.}\label{fig_sol}
\end{figure}

By using system~\eqref{sys_ini} under the SVEA (this is analogous to what is done in Ref.~\onlinecite{talanina} and valid only out of the gap, where $k'\rightarrow\omega'$ and $\sigma_2$ tends to become the dominant nonlinear coefficient in Eq.~\eqref{third}):
\begin{equation}\label{sys_svea}
\left\{\begin{array}{l}
 \partial_x\psi=i\Gamma_\rfm\lambda\varphi/2,\\
 \partial_\tau\varphi=i\left(\left[i\gamma'_\rmx+\Delta\omega'_\rmx-|\varphi|^2\right]\varphi+\Gamma_\rmf\psi\right),\\
      \end{array}\right.
\end{equation}
where $\tau=T-x$, we simulated the propagation dynamics of a short intense sech-shaped pulse in the excitonic medium and demonstrated solitary waves formation for the out-of-gap regime (see Fig.~\ref{fig_sol}(a)) and an essential broadening of the spectrum (see Fig.~\ref{fig_sol}(b)). In the case of $\omega\gtrsim\omega_\rmx$ this spectral broadening, as well as the MI sidebands detuning from $\omega$, should be limited by the exciton binding frequency value to avoid the excitation of band electrons. Here we employed a numerical method similar to that used in Sec.~\ref{s3}.

\section{Realistic excitonic waveguide parameters\label{s5}}
On the basis of above results we can discuss now the structure and parameters of a realistic excitonic waveguide and the physical conditions (and constraints) for the observation of considered effects by the example of some specific excitonic material.

As it was mentioned previously, GaAs was selected as the representative semiconducting excitonic material, for which at low temperatures and for high-quality samples:~\cite{hobden,tredi,schaefer} $\hbar\omega_\rmx=1515$~meV, $\hbar\omega_\mathrm{b}=4$~meV, $\hbar\tilde{a}=0.08$~meV, $\hbar\gamma_\rmx\simeq0.03$~meV, and $\eb=12.56$. The low temperature condition is mentioned here because, as it was shown experimentally,~\cite{tredi,schaefer} for GaAs the polaritonic effects take place only below $\sim20~\mathrm{K}$. However, for those samples and temperatures, for which $\hbar\tilde{a}<\hbar\gamma_\rmx$ (for example, in presence of inhomogeneous broadening), the polaritonic gap will not be visible and only out-of-gap effects can be observed.

According to Refs.~\onlinecite{knorr2,kling} one can properly return dimensionality to the macroscopic polarization field in the following way: $P\rightarrow uP$, where $u=[2d_\mathrm{cv}/(\pi a_0^3\ee)][\ee cn/2]^{1/2}$ ($P$ is measured in the units of intensity~\cite{agrawal}), $d_\mathrm{cv}$ is the interband dipole matrix element, $a_0$ is the 1s-exciton Bohr radius, $\ee$ is the vacuum permittivity, and the spin summation is also taken into account. Therefore, the nonlinear coefficients in Eq.~\eqref{polar_ini} take form $\{\alpha;\beta\}\rightarrow\{\alpha;\beta\}/u^2$, while the coupling parameter becomes equal~\cite{ostr} $\tilde{a}=2d^2_\mathrm{cv}/(\pi a_0^3 \hbar\ee\eb)$ in complete accordance with the polaritonic gap definition.~\cite{haug} For GaAs, $a_0\simeq12$~nm and $d_\mathrm{cv}/e\simeq0.4$~nm. These parameters allow one to calculate the corresponding nonlinear coefficients: $\alpha\simeq2040~\mathrm{m}^2/(\mathrm{W~s})$ and $\beta\simeq2.33\cdot10^{-10}~\mathrm{m}^2/\mathrm{W}$. Now we can estimate the limiting field intensities from the condition $|P|^2\sim1/\beta$: $P^2_\mathrm{lim}\simeq0.43~\mathrm{MW/cm^2}$ and $E^2_\mathrm{lim}\simeq[\alpha P^3_\mathrm{lim}/(\tilde{a}\eb)]^2\simeq14.3~\mathrm{MW/cm^2}$. If the incident pulse intensity is well below this limiting one the density of excitons in GaAs waveguide is small enough and both Eq.~\eqref{polar_main} and the developed theory as a whole remain valid. It should be noted here that the damage threshold for the chosen material is even higher than $E^2_\mathrm{lim}$: about $40~\mathrm{MW/cm^2}$ for a 20~ns incident pulse.~\cite{smith} As this threshold grows rapidly with the reduction of incident pulse duration~\cite{smith} one can expect that it is much higher for ps-pulses considered in this work. Knowing the nonlinear coefficient $\alpha$ one can also easily reproduce the input pulse intensities, which are necessary to form solitons in the excitonic waveguide. According to the used scaling of electric field ($E_0=1/(\sqrt{\alpha t_0}\tilde{a}t_0\eb)$), for example, for solitons with a temporal width $t_0=2/\tilde{a}$ (see Fig.~\ref{fig_ampl}) these intensities are: $\sim10-100~\mathrm{kW/cm^2}$ for incident pulses spectrally centered inside the polaritonic gap and $\sim0.1-1~\mathrm{kW/cm^2}$ for those spectrally centered outside the gap (it is clear that in both cases these intensities are well below the limiting ones).

Nonlinear effects such as solitons formation could be observed, for example, in silica optical fibers possessing a Kerr nonlinearity only at much higher pulse intensities~\cite{agrawal} ($\lesssim 1~\mathrm{MW/cm^2}$). The source of this large difference in intensities can be found by the proper comparison of the nonlinear coefficient $\alpha$ at the cubic term in Eq.~\eqref{polar_main} with the corresponding coefficient of the Kerr effect in silica fibers~\cite{agrawal} (which is actually the nonlinear refractive index $n^\mathrm{SiO_2}_2\simeq10^{-20}~\mathrm{m^2/W}$). As we will see below the excitonic nonlinearity is much stronger than the Kerr one, however it has a resonant nature - it decreases rapidly with the increase of detuning. This can be demonstrated by the example of plane wave propagation. In this case, according to Eq.~\eqref{polar_main} the amplitudes of polarization and electric field are related by the following expansion up to the third order:
\[
P\approx-\frac{\tilde{a}\eb \Gamma_\rmf}{\Delta\omega_\rmx+i\gamma_\rmx}E-\alpha_\mathrm{eff} E^3,\quad \alpha_\mathrm{eff}\equiv\frac{(\tilde{a}\eb \Gamma_\rmf)^3 \Gamma_1}{(\Delta\omega_\rmx+i\gamma_\rmx)^4}\alpha.
\]
This dependence clearly shows that the factor multiplying the cubic nonlinear term decreases rapidly when the detuning $\Delta\omega_\rmx$ grows. In other words, if the incident pulse is spectrally centered far enough from the exciton resonance the considered nonlinearity becomes negligible, and the background Kerr nonlinearity of the medium starts to play the main role. However, near the resonance ($\Delta\omega_\rmx\sim10\tilde{a}$) for reasonable~\cite{schaefer} low-temperature values of $\hbar\gamma_\rmx\simeq0.03-0.3$~meV one can estimate $\re\alpha_\mathrm{eff}\sim10^{11}-10^{10}n^\mathrm{SiO_2}_2$, that clearly indicates the domination of excitonic nonlinearity in the vicinity of resonance.

\begin{figure}[t]
  \includegraphics[width=6cm]{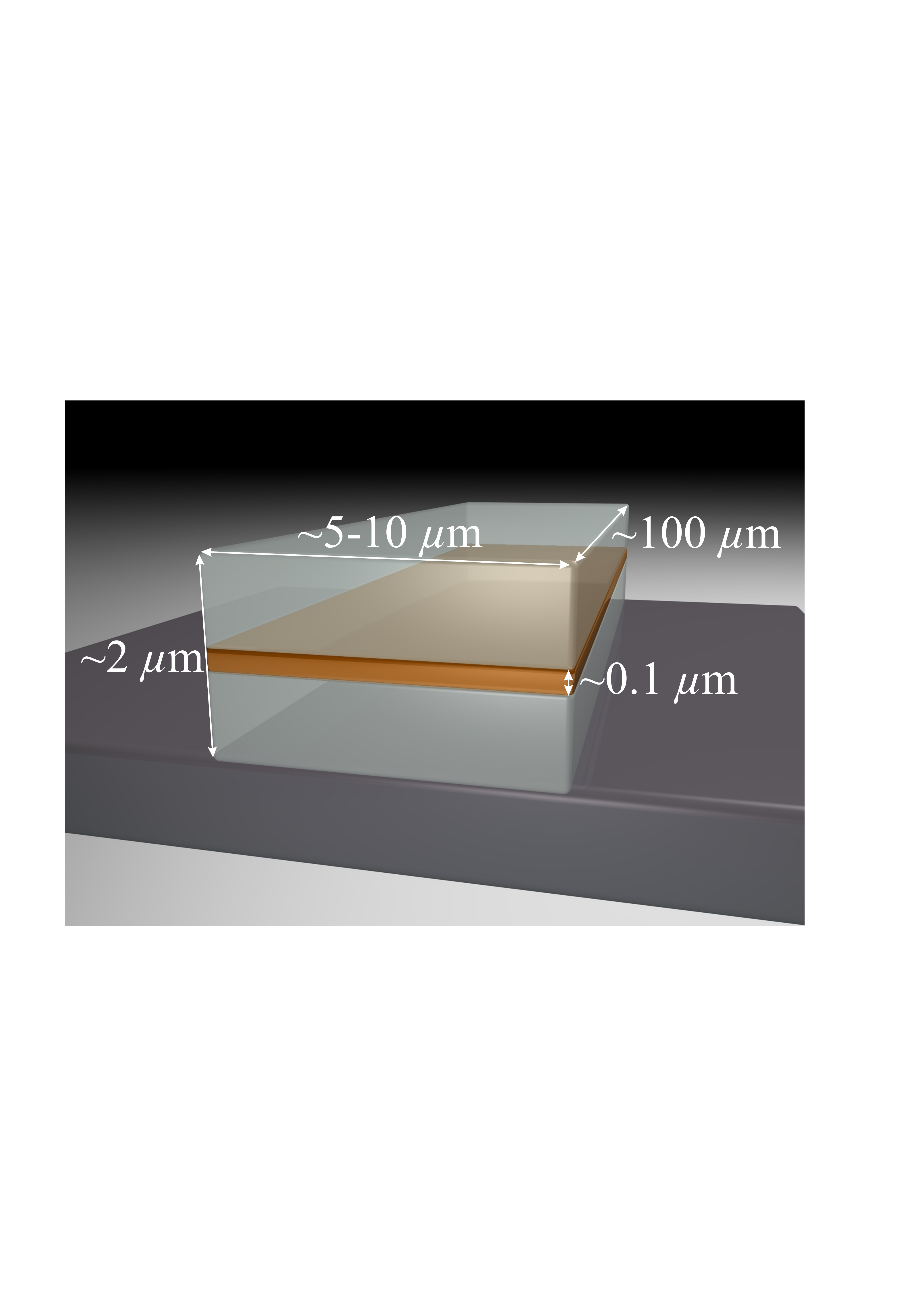}
  \caption{Schematic drawing of a sandwiched semiconducting waveguide. The realistic length parameters are indicated for the case of GaAs (central layer) surrounded by AlGaAs.}\label{fig_semi}
\end{figure}

It is also important to note that excitonic media efficiently convert light into excitons, thus being optically highly absorptive near the exciton resonance. Therefore, the incident pulse spectrum central frequency should not be too close to $\omega_\rmx$. One can estimate the corresponding absorption length as~\cite{haug}
\begin{equation}\label{abs}
l=\frac{c}{\omega}\frac{\re{\sqrt{\varepsilon}}}{\im{\varepsilon}},\quad \varepsilon=\eb\left(1-\frac{\tilde{a}\Gamma_\rmf \Gamma_\rfm}{\Delta\omega_\rmx+i\gamma_\rmx}\right),
\end{equation}
where $\varepsilon$ is the polariton dielectric function in the linear regime. In the case of GaAs, for reasonable values of the detuning (a few times $\tilde{a}$), this length is of the order of several microns. This agrees well with the micron-scale lengths, which are completely sufficient for the revelation of nonlinear optical effects in excitonic waveguides (see Sec.~\ref{s3}). However, waveguide lengths of just a few microns are not sufficient to actually guide light, and it is unlikely that one can produce such a short waveguide made of pure semiconducting material. One possible solution is to use a composite waveguide made of material without excitonic resonances in its optical response (e.g., silica) doped with macroscopic domains of semiconductor (for example, nanoparticles). But then a different theory based on the Maxwell-Garnett formalism should be developed. Another way is to use a waveguide in the form of sandwiched semiconducting heterostructure with central working layer made of semiconductor with an excitonic feature surrounded by the semiconducting layers made of a different material (see Fig.~\ref{fig_semi}). This cladding material should have no excitonic resonances in the same range of frequencies than that in the working layer so that the propagating light will not be affected by the excitonic nonlinearity, but should possess almost the same refractive index so that the optical guided mode is well spread outside the working layer, and the photon-exciton coupling is reduced accordingly.

Because of this special structure of the proposed waveguide, the modal distribution $F$ for the electric field will be much wider than the distribution $M$ for the polarization field. Therefore, the transverse integral $\Gamma_\rfm$ in the main system~\eqref{sys_ini} becomes much smaller than unity, while still $\Gamma_\rmf\gtrsim1$. For example, by using the realistic AlGaAs/GaAs sandwiched waveguide transverse parameters, which are indicated at Fig.~\ref{fig_semi}, one can obtain $\Gamma_\rfm\ll1$ and then get from Eq.~\eqref{abs} for $\Delta\omega_\rmx\sim10\tilde{a}$ an absorption length $l\sim100~\mu\mathrm{m}$. Of course, as it follows from system~\eqref{sys_ini} such a small $\Gamma_\rfm$ leads to a weak coupling between the electric and polarization fields and thus to a weak influence of the excitonic nonlinearity on the electric field, but this is compensated by the long propagation distance. Also the reduction of $\Gamma_\rfm$ results in decrease of the gap width $\lambda_\Gamma$ (see Sec.~\ref{s4}), therefore longer incident pulses should be used to have a sufficiently narrow spectrum and efficiently deliver the pulse energy into the polaritonic gap.

As a last point we should also note that low-dimensional semiconductor structures, such as semiconductor microcavities,~\cite{dmitry} certainly have a number of advantages with respect to bulk semiconductors considered here (e.g., larger exciton binding frequency, which translates into larger optical nonlinearities). However, waveguides based on a bulk semiconductor are preferable for observation of nonlinear processes associated with solitons because they are weakly affected by diffraction, which is hard to overcome in microcavities,~\cite{dmitry} and also because such waveguides can be more easily produced and tested than guiding structures in microcavities. A possibility to properly combine the excellent waveguiding properties of an optical fiber with optical features of pure bulk semiconductors in a single structure has been very recently demonstrated - a silica optical fiber with a semiconducting ZnSe core has been fabricated.~\cite{sparks} This should also stimulate further experimental investigations of nonlinear optical properties of bulk semiconductor waveguides.

\section{Conclusions\label{s6}}
In conclusion, we have carried out a MI analysis of continuous waves propagation in semiconducting excitonic media. Strong peaks of MI relatively far-detuned from the pump frequency have been found, possessing gain orders of magnitude larger than that in the case of optical fibers. Our analytical findings are supported by numerical simulations of wide super-Gaussian pulses propagation. Moreover, the solitonic sector shows evidence of out-of-gap and in-gap solitons, the shape of which has been given analytically by using SVEA in a unified formalism, greatly generalizing previous theories. Last but not least, we have proposed a simple way to overcome the problem of light-exciton conversion absorption, by employing a sandwiched excitonic structure (by the example of AlGaAs/GaAs). Further work will consider the complete set of semiconductor Bloch equations in a semiclassical way (including the equation for inversion) instead of Eq.~\eqref{polar_ini}. This will also allow to treat the evolution of short (sub-picosecond) and coherent pulses beyond the approximation of low exciton density.

\begin{acknowledgments}
This work was supported by the German Max Planck Society for the Advancement of Science (MPG).
\end{acknowledgments}


\begin{thebibliography}{99}

\bibitem{agrawal} G. P. Agrawal, {\it Nonlinear Fiber Optics}, 4th ed. (Academic Press, San Diego, 2007).

\bibitem{wadsworth} W. J. Wadsworth, N. Joly, J. C. Knight, T. A. Birks, F. Biancalana, and P. St. J. Russell, Opt. Express {\bf 12}, 299 (2004).

\bibitem{hasegawa} A. Hasegawa and F. Tappert, Appl. Phys. Lett. {\bf 23}, 142 (1973).

\bibitem{mollenauer} L. F. Mollenauer and J. P. Gordon, {\it Solitons in Optical Fibers} (Elsevier, 2006).

\bibitem{dudley} J. M. Dudley, G. Genty, and S. Coen, Rev. Mod. Phys. {\bf 78}, 1135 (2006).

\bibitem{akimoto} O. Akimoto and K. Ikeda, J. Phys. A: Math. Gen. {\bf 10}, 425 (1977); K. Ikeda and O. Akimoto, \ib \textbf{12}, 1105 (1979).

\bibitem{knorr} A. Knorr, R. Binder, M. Lindberg, and S. W. Koch, Phys. Rev. A {\bf 46}, 7179 (1992).

\bibitem{ostr} Th. \"{O}streich and A. Knorr, Phys. Rev. B {\bf 50}, 5717 (1994).

\bibitem{knorr2} A. Knorr, Th. \"{O}streich, K. Sch\"{o}nhammer, R. Binder, and S. W. Koch, Phys. Rev. B {\bf 49}, 14024 (1994).

\bibitem{talanina93} I. B. Talanina, M. A. Collins, and V. M. Agranovich, Solid State Commun. {\bf 88}, 541 (1993); Phys. Rev. B {\bf 49}, R1517 (1994); I. B. Talanina, Solid State Commun. {\bf 97}, 273 (1996).

\bibitem{talanina} I. Talanina, D. Burak, R. Binder, H. Giessen, and N. Peyghambarian, Phys. Rev. E {\bf 58}, 1074 (1998).

\bibitem{giessen} H. Giessen, A. Knorr, S. Haas, S. W. Koch, S. Linden, J. Kuhl, M. Hetterich, M. Gr\"{u}n, and C. Klingshirn, Phys. Rev. Lett. {\bf 81}, 4260 (1998); N. C. Nielsen, T. H\"{o}ner zu Siederdissen, J. Kuhl, M. Schaarschmidt, J. F\"{o}rstner, A. Knorr, and H. Giessen, \ib {\bf 94}, 057406 (2005); T. H\"{o}ner zu Siederdissen, N. C. Nielsen, J. Kuhl, M. Schaarschmidt, J. F\"{o}rstner, A. Knorr, G. Khitrova, H. Gibbs, S. W. Koch, and H. Giessen, Opt. Lett. {\bf 30}, 1384 (2005).

\bibitem{kamch} S. A. Darmanyan, A. M. Kamchatnov, and M. Nevi\'ere, JETP {\bf 96}, 876 (2003); A. M. Kamchatnov, S. A. Darmanyan, and M. Nevi\'ere, J. Luminesc. {\bf 110}, 373 (2004).

\bibitem{dmitry} O. A. Egorov, D. V. Skryabin, A. V. Yulin, and F. Lederer, Phys. Rev. Lett. {\bf 102}, 153904 (2009); O. A. Egorov, A. V. Gorbach, F. Lederer, and D. V. Skryabin, \ib {\bf 105}, 073903 (2010).

\bibitem{bianex}  F. Biancalana, S. B. Healy, R. Fehse, and E. P. O'Reilly, Phys. Rev. A {\bf 73}, 063826 (2006); F. Biancalana and C. Creatore, Opt. Express {\bf 16}, 14882 (2008); F. Biancalana, L. Mouchliadis, C. Creatore, S. Osborne, and W. Langbein, Phys. Rev. B {\bf 80}, 121306(R) (2009).

\bibitem{stoychev} K. T. Stoychev and M. T. Primatarova, J. Phys.: Condens. Matter {\bf 13}, L183 (2001).

\bibitem{mihalache} V. E. Zakharov and A. B. Shabat, Sov. Phys. JETP {\bf 34}, 62 (1972); D. Mihalache, N. Truta, N. C. Panoiu, and D.-M. Baboiu, Phys. Rev. A {\bf 47}, 3190 (1993).

\bibitem{hobden} M. V. Hobden and M. D. Sturge, Proc. Phys. Soc. {\bf 78}, 615 (1961).

\bibitem{tredi} A. Tredicucci, Y. Chen, F. Bassani, J. Massies, C. Deparis, and G. Neu, Phys. Rev. B {\bf 47}, 10348 (1993).

\bibitem{schaefer} A. C. Schaefer and D. G. Steel, Phys. Rev. Lett. {\bf 79}, 4870 (1997).

\bibitem{kling} C. F. Klingshirn, {\it Semiconductor Optics}, 3rd ed. (Springer-Verlag, Berlin Heidelberg, 2007).

\bibitem{haug} H. Haug and S. W. Koch, {\it Quantum Theory of the Optical and Electronic Properties of Semiconductors}, 4th ed. (World Scientific, Singapore, 2005).

\bibitem{smith} J. L. Smith and G. A. Tanton, Appl. Phys. {\bf 4}, 313-315 (1974).

\bibitem{sparks} J. R. Sparks, R. He, N. Healy, M. Krishnamurthi, A. C. Peacock, P. J. A. Sazio, V. Gopalan, and J. V. Badding, Adv. Mater. {\bf 23}, 1647 (2011).
\end{thebibliography}
\end{document}